# Image Steganography based on a Parameterized Canny Edge Detection Algorithm

Youssef Bassil
LACSC – Lebanese Association for Computational Sciences
Registered under No. 957, 2011, Beirut, Lebanon

## ABSTRACT
Steganography is the science of hiding digital information in such a way that no one can suspect its existence. Unlike cryptography which may arouse suspicions, steganography is a stealthy method that enables data communication in total secrecy. Steganography has many requirements, the foremost one is irrecoverability which refers to how hard it is for someone apart from the original communicating parties to detect and recover the hidden data out of the secret communication. A good strategy to guaranteeirrecoverability is to cover the secret data not usinga trivial method based on a predictable algorithm, but using a specific random pattern based on a mathematical algorithm. This paper proposes an image steganography technique based on theCanny edge detection algorithm.It is designed to hide secret data into a digital image within the pixels that make up the boundaries of objects detected in the image. More specifically, bits of the secret data replace the three LSBs of every color channel of the pixels detected by the Canny edge detection algorithm as part of the edges in the carrier image. Besides, the algorithm is parameterized by three parameters: The size of the Gaussian filter, a low threshold value, and a high threshold value. These parameters can yield to different outputs for the same input image and secret data. As a result, discovering the inner-workings of the algorithm would be considerably ambiguous, misguiding steganalysts from the exact location of the covert data. Experiments showed a simulation tool codenamed GhostBit, meant to cover and uncover secret data using the proposed algorithm. As future work, examining how other image processing techniques such as brightness and contrast adjustment can be taken advantage of in steganography with the purpose ofgiving the communicating parties more preferences tomanipulate their secret communication.

## 1. INTRODUCTION
Steganography is the science of hiding data into another form of data, for instance, hiding a plaintext message into an image file [1]. Steganography has been exploited throughout history by individuals, military, secret intelligence, and governments to stealthily communicate and transmit information without drawing any attraction. It has numerous applications which range from secret communication, to digital watermarking, data integrity, copyright protection, and data tampering [2]. Cryptography is yet another security mechanism that existed years before steganography. Unlike steganography which hides the secret message into an innocent-looking carrier file to avoid being detected, cryptography converts the secret message into a scrambled form that is unreadable by third parties. In other words, cryptography ciphersthe secret message so that it cannot be read by eavesdroppers; while, steganography conceals the secret message so that it cannot be seen by eavesdroppers [3]. One of steganography requirements is robustness denoting how much a carrier file can withstand image processing operations such as transformation, compression, and cropping. Another requirement is the hiding capacity which indicates the amount of secret data that can be embedded into a particular carrier file [4]. Even though these two aforementioned requirements are critical, steganography is of no avail without two last requirements, mainly imperceptibility and irrecoverability. In fact, the imperceptibility of steganography comes from the weakness of the human visual system (HVS) which cannot perceive the slight variation of colors at the high frequency side of the visual spectrum. For instance, the difference between these two colors $11111110_2$ and $11111111_2$ is impossible to be detected by the human eye [5]. In effect, the imperceptibility of steganography ensures that no one apart from the sender and receiver knows about the fact that a secret message is being communicated. On the other hand, the irrecoverabilityof steganography refers to how hard it is for someone apart from the original sender and the intended receiver to detect and recover the covertsecret data out of the carrier file. A good approach to ensure irrecoverability is to cover the secret data not in every pixel of the carrier image but using a specific pattern that is governed by some mathematical formula or algorithm. For instance, as edge detection algorithms can identify object boundaries in the image, they can be used to select part of the pixels of the carrier image to hold the secret data. Furthermore, little has been done to exploit the properties of edge detection in the area of information hiding and image steganography.

This paper proposes a new image steganography technique based on Canny Edge Detection algorithm. It conceals secret data, regardless of their type, into the pixels that compose the boundaries of objects detected in the carrier image. Technically speaking, bits of the secret data substitutethe three LSBs of every color channel of the pixels detected by the Canny edge detection algorithm as part of the edges in the carrier image. Additionally, the Canny edge detection algorithm can be parameterized by the communicating parties using three parameters: The size of the Gaussian filter, a low threshold value and a high threshold value. These parameters change the way the edge detection algorithm detects edges in the image. It is by parameterizing the steganography algorithm, different carrier pixels can be obtained for the same input image and secret data. As a result, identifying how the algorithm works would be quite complicated and ambiguous for eavesdroppers, misguiding them from the actual location of the covert data.

## 2. STEGANOGRAPHY & ITS PROPERTIES

Steganography refers to concealing data in an overt carrier file in such a way that it is difficult for unauthorized third parties to detect and recover the hidden data. As a result, the effectiveness of a steganography algorithm is determined by three properties: The capacity of data that can hidden without distorting the carrier file; the imperceptibility of the carrier file after hiding the secret data into it; and the irrecoverabilityof the hidden data in case they were detected [6, 7].

**Hiding Capacity:** It determines the number of bytes that can be covered within the carrier file without distorting or damaging it. For instance, in image steganography, it is important to hide as much as possible data inside the carrier image without increasing its brightness, without making it blurry, without pixelizing it, and without changing its size. This would be a key element in making the hidden data imperceptible and the carrier image innocent and unsuspicious.

**Imperceptibility:**It refers to the ability of the steganography algorithm to hide data in an undetectable way so much so that no one can see any visible artifacts or distortions in the carrier file. It therefore avoids drawing suspicions and obscures the fact that a secret communication is taking place.

**Irrecoverability:** It refers to how much an intercepted carrier file can be easily decoded and reverse-engineered so as to extract the data hidden inside it. An irrecoverable steganography algorithm makes it hard for eavesdroppers and unauthorized third parties to recover the hidden data from the carrier file despite knowing that steganography has been employed.

## 3. STATE-OF-THE-ART IN IMAGE STEGANOGRAPHY

So far, massive research work has been conducted in the development of steganography for digital images. One of the earliest techniques is the LSB technique which obscures data communication by inserting the secret data into the insignificant parts of the pixels of an image file, more particularly, into the least significant bits (LSB) [8]. The modified version of the image, which is called carrier file or stego file, is then sent to the receiver through a public channel. The foremost requirement of the LSB technique is that it should not exhibit any visual signs in the carrier image so as to not give any indications that secret data are being communicated covertly.

Basically, the LSB technique is an insertion-based image steganography method that embeds secret data into uncompressed computer image files such as BMP and TIFF. In this technique, the data to hide are first converted into a series of bytes, then into a series of smaller chunks each of which is of size $n$ bits. Then, $n$ LSBs of the pixels of the carrier image are replaced by each of the chunks of the original data to hide. The ultimate result of this operation is a carrier image carrying the secret data into the LSBs of its pixels. As the color values that are determined by LSBs are insignificant to the naked eye, it is hard to tell the difference between the original image and the tampered one, taking into consideration that no more than a certain number of LSBs were used to conceal the secret data; Otherwise, visual artifacts and damages would be produced in the carrier image which would in turn draw suspicions and raise attention about something unusual in the carrier image. For instance, in 24-bit True Color BMP images, using more than three LSBs per color component to hide data may result in perceptible artifacts in the carrier image [9]. As an illustration for the LSB technique, let's say that the letter H needs to be hidden into an 8-bit grayscale bitmap image. The ASCII representation for letter H is 72 in decimal or 01001000 in binary. Assuming that the letter H is divided into four chunks each of 2 bits, then four pixels are needed to totally hide the letter H. Moreover, assuming that four consecutive pixels are selected from the original image whose grayscale values are denoted by $P_1$=11011000, $P_2$=00110110, $P_3$=11001111, and $P_4$=10100011, then substituting every two LSBs in every of these four pixels by a 2-bit chunk of the letter H, would result in a new set of pixels denoted by $P_1$= 110110**01**, $P_2$=001101**00**, $P_3$=110011**10**, and $P_4$=101000**00**. Despite changing the actual grayscale values of the pixels, this has little impact on the visual appearance of the carrier image because characteristically, the Human Visual System (HVS) cannot differentiate between two images whose color values in the high frequency spectrum are marginally unalike [10].

On the other hand, other steganography techniques and algorithms for digital images have been proposed and researched both in spatial and frequency domains. They include masking and filtering [11], encrypt and scatter [12], transformation [13], and BPCS [14] techniques.

**Masking and Filtering Technique:** This technique is based on digital watermarking but instead of increasing too much the luminance of the masked area to create the digital watermark, a small increase of luminance is applied to the masked area making it unnoticeable and undetected by the naked eye. As a result, the lesser the luminance alteration, little the chance the secret message can be detected. Masking and filtering technique embeds data in significant areas of the image so that the concealed message is more integral to the carrier file.

**Encrypt and Scatter Technique:** This technique attempts to emulate what is known by White Noise Storm which is a combination of spread spectrum and frequency hopping practices. Its principle is so simple; it scatters the message to hide over an image within a random number defined by a window size and several data channels. It uses eight channels each of which represents 1 bit; and consequently, each image window can hold 1 byte of data and a set of other useless bits. These channels can perform bit permutation using rotation and swapping operations such as rotating 1 bit to the left or swapping the bit in position 3 with the bit in position 6. The niche of this approach is that even if the bits are extracted, they will look garbage unless the permutation algorithm is first discovered. Additionally, the encrypt and scatter technique employs DES encryption to cipher the message before being scattered and hidden in the carrier file.

**Transformation Technique**: This technique is often used in the lossy compression domain, for instance, with JPG digital images. In fact, JPG images use the discrete cosine transform (DCT) to perform compression. As the cosine values cannot be calculated accurately, the DCT yields to a lossy compression. The transformation-based steganography algorithms first compress the secret message to hide using DCT and then integrate it within the JPG image. That way, the secret message would be integral to the image and would be hard to be decoded unless the image is first decompressed and the location of the hidden message is recovered.

**BPCS Technique**:This technique which stands for Bit-Plane Complexity Segmentation Steganography, is based on a special characteristic of the Human Visual System (HVS). Basically, the HVS cannot perceive a too complicated visual pattern as a coherent shape. For example, on a flat homogenous wooden pavement, all floor tiles look the same. They visually just appear as a paved wooden surface, without any indication of shape. However, if someone looks closely, every collection of tiles exhibits different shapes due to the particles that make up the wooden tile. Such types of images are called vessel images. BPCS Steganography makes use of this characteristic by substituting complex regions on the bit-planes of a particular vessel image with data patterns from the secret data.

## 4. PROPOSED TECHNIQUE

This paper proposes a new image steganography technique based on Canny Edge Detection algorithm. It hides secret data into the pixels that make up the extracted edges of the carrier image. The secret data can be of any type, not necessarily text, and they are actually concealed into the three LSBs (Least Significant Bits) of the pixels of the carrier image, but not in every pixel, only in the ones that are part of the edges detected by the Canny edge detection algorithm. The proposed technique first reads a 24-bit uncompressed-type colored input image and transforms it into a monochrome version making its bit-depth equal to 1, i.e., all pixels are either black or white. The black pixels are represented by bit 0; while, the white pixels are represented by bit 1. Then, the Canny edge detection algorithm is applied on the transformed image to extract the edges from it which would be revealed as white lines and curves (i.e. composed of only white pixels whose values are 1). Consequently, bits of the secret data are stored in the three LSBs of every color channel of the pixels that make up the extracted edges in the original 24-bit colored input image. In other words, the secret data would be hidden only in the pixels of the original image that were indicated by the Canny edge detection algorithm as part of the detected edges. The resulting image is a 24-bit image called carrier image housing the secret data into the pixels of its edges denoted by the Canny algorithm. Furthermore, the Canny edge detection algorithm can be parameterized by the communicating parties (i.e., sender and receiver) using three parameters: The size of the Gaussian filter, a low threshold value and a high threshold value. These parameters change the effectiveness of the algorithm; and thus, either more or fewer edges can be detected in the image. These parameters are stored in a predefined location in the carrier image that is known to the communicating parties. Same parameters must be used for the covering and the uncovering processes or else the secret data cannot be recovered. In fact, in order to recover the secret data, the carrier image has first to be converted into a monochrome version. Then, the Canny edge detection algorithm has to be applied on the converted image to reveal the pixels of the edges. In effect, these pixels are the actual carrier pixels into which the secret data are hidden in the three LSBs of their every color channel.

### 4.1 Canny Edge Detection

Fundamentally, edge detection is the process of identifying points in a computer image at which the image brightness changes abruptly, for instance, pixels deviating from low intensities to high intensities or vice versa, exhibiting some discontinuities [15]. The aim of edge detection is to detect and capture important events and changes in the properties of the image. These changes are due to the following events: Discontinuities in depth, discontinuities in surface orientation, changes in material properties, and changes in scene illumination [16, 17]. Often, applying edge detection on a digital image may result into a set of linked lines and curves that indicate the boundaries of objects in the image. The advantage of edge detection is that it reduces the amount of data to be analyzed and processed by image processing algorithms. It in fact removes the less relevant information while preserving the more significant structural features of the image. In practice, many edge detection algorithms have been used, however, one of the most efficient and widely known algorithm [18] is the Canny edge detection which was developed by John Canny in 1986 [19]. In essence, the Canny edge detection is a multi-stage algorithm that can extract a wide range of edges in images regardless of the noise present in them. The Canny edge detection algorithm has the following five stages:

1. Smoothing: Blur the image to eliminate noise.
2. Searching for gradients: Find the edge strength in the image by taking the gradient with large magnitudes.
3. Non-maximum suppression: Mark local maxima as edges.
4. Double thresholding: Find possible edges by computing thresholding.
5. Edge linking: Final edges are found by discarding all edges that are not connected to strong edges.

Moreover, the Canny algorithm has two basic adjustable parameters, the size of the Gaussian filter and the threshold.

The size of the Gaussian filter: As Gaussian filter is used in the first stage to smooth the image and reduce the noise and unwanted details, decreasing its size would result in blurry image that allows the detection of smaller details. In contrast, increasing its size would result in even more blurry images, spreading the value of a given pixel over a larger area of the image and allowing the detection of larger edges.

Thresholds: The Canny algorithm uses two thresholds which allow more flexibility for edge detection. Often, a threshold set too high can miss important information; whereas, a threshold set too low can extract irrelevant information such as noise.

Figure 1 depicts the input (a) and the output (b) of a Canny edge detection algorithm with Gaussian filter set to 2, low threshold set to 20, and high threshold set to 30.

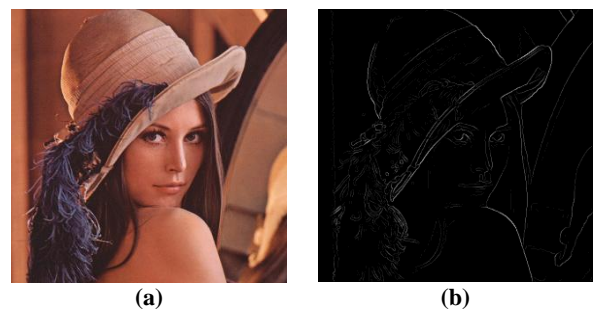

(a)          (b)

**Fig 1: Input (a) and Output (b) of Canny Edge Detection Algorithm**

### 4.2 Proposed Algorithm

The proposed algorithm is used to perform two processes: The covering process which hides secret data into a digital image turning it into a carrier image, and the uncovering process which recovers secret data out of the carrier image. The covering process has several steps to be executed and they are as follows:

1. The algorithm reads a 24-bit uncompressed image, possibly colored, denoted by $I=\{p_1,p_2,p_3,p_{n-1}\}$ where $p_i$ is the $i^{th}$ pixel in the image and $n$ is the total number of pixels. In the same context, every pixel $p_i$ can be represented as $p_i=\{$ $R_i$ $[r_0,r_1,r_2,r_3,r_4,$ $r_5,r_6,r_7]$, $G_i$ $[g_0,g_1,g_2,g_3,g_4,$ $g_5,g_6,g_7]$, $B_i$ $[b_0,b_1,b_2,b_3,b_4,b_5,b_6,b_7]$ $\}$ where $i$ is the index of the $i^{th}$ pixel, $r_j$ is the $j^{th}$ bit of color channel R, $g_j$ is the $j^{th}$ bit of color channel G, and $b_j$ is the $j^{th}$ bit of color channel B. It is worth noting that because $I$ is a 24-bit image, each of its pixels is made up of three color channels each of which is of length 8 bits.

2. The algorithm converts $I$ into a monochrome version denoted by $f(I)=I'$ where $I'$ is no more composed of 24-bit pixels each containing three color channels; rather, it is composed of single-bit pixels whose value can be either 0 representing the black color or 1 representing the white color. The purpose of this conversion is to ease the processing of subsequent steps.

3. Three parameters for the Canny edge detection algorithm are defined by the communicating parties. They include 1) the size of the Gaussian filter whose value can range between 1.0 and 3.0, 2) a low threshold whose value can range from 0 to 255, and 3) a high threshold whose value can range from 0 to 255.

4. The Candy edge detection algorithm is executed on $I'$ using the three parameters selected in step 3. The results are a collection of lines, curves, and points denoting the edges or the boundaries of the objects in the image $I'$. The pixels that constitute the extracted edges are denoted by $E=\{e_1,e_2,e_3,e_{m-1}\}$ where $e_k$ is the $k^{th}$ pixel that makes up the edges and $m$ is the total number of these pixels.

5. The algorithm reads the secret data to hide denoted by $D=\{d_0,d_1,d_2,...d_{t-1}\}$ where $d$ is a single bit in D and $t$ is the total number of bits in D.

6. Every three LSBs of every color channel of pixels $p_k$ in $I$ are substituted by three bits of the secret data D. It is worth noting that $p_k$ is a pixel in the original image $I$ whose index is pointed to by $k$ which is a pixel in E that is part of the detected edges.

7. Step 6 is repeated until all bits in D are exhausted.

8. The obtained image is visually identical to $I$ but with the three LSBs ofits every color channel replaced by three bits of the secret data D. It is called the carrier image and it is denoted by $C=\{p_1,p_2,p_3,p_{n-1}\}$ where $p_k=\{$ $R_k$ $[r_0,r_1,r_2,r_3,r_4,$ $d_q,d_{q+1},d_{q+2}]$, $G_k$ $[g_0,g_1,g_2,g_3,g_4,$ $d_{q+3},d_{q+4},d_{q+5}]$, $B_k$ $[b_0,b_1,b_2,b_3,b_4,$ $d_{q+6},d_{q+7},d_{q+8}]$ $\}$

9. The three parameters namely the size of the Gaussian filter, the low threshold, and the high threshold that were already specified by the sender have to be communicated with the receiver prior to starting the secret communication. In effect, many solutions are possible, one of which is sending them via email, or handing them over the phone, or injecting them at the end of the carrier image, or embedding them into some predefined pixels locations in the carrier image using the traditional LSB technique.

10. Finally, the carrier image C is sent to the receiver.

As for the uncovering process, it is the reverse of the above steps. First, the carrier image is received by the receiver. Then, it is converted into a monochrome version, and then the Canny edge detection algorithm is applied on it. The locations of the pixels of the resulting edges are used to point to the carrier pixels in the carrier image. The three LSBs of every color channel of these carrier pixels are extracted one after the other. Eventually, their concatenation yields to the secret data. Figure 2 depicts the flowchart of the proposed algorithm illustrating how the covering process works.

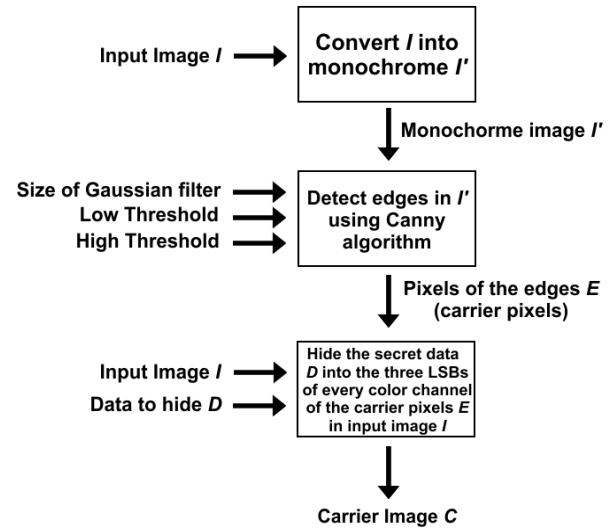

**Fig 2: Flowchart of the Proposed Algorithm**

## 5. EXPERIMENTS AND RESULTS

For experimentation purposes, a simulation software was built using MS Visual C# 4.0 and MS Visual Studio 2012 under the MS .Net Framework 4.0 [20]. The software is codenamed GhostBit and it is capable of covering and uncovering secret data using the proposed method and algorithm. The Canny edge detection algorithm was implementation using a third party library called AForge.Net framework [21]. It is actually a collection of application programming interfaces designed for developers and researchers in the fields of Computer Vision and Artificial Intelligence including image processing, computer vision, neural networks, etc. Figure 3 shows the main GUI interface of GhostBit.

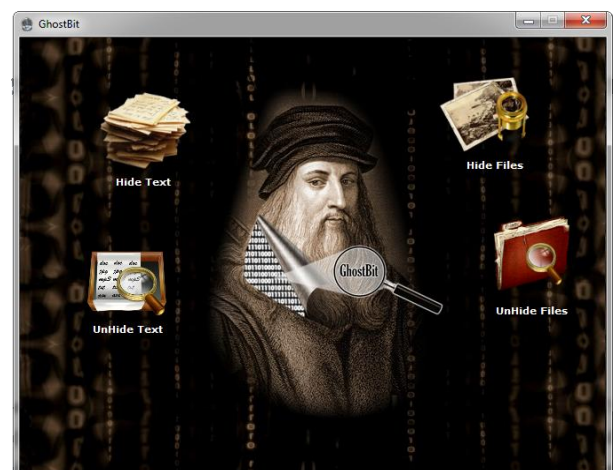

**Fig 3:GhostBit Main GUI**

The input image is a 24-bit BMP image depicted in Figure 4. Figure 5 shows the results of applying the Canny edge detection algorithm on the input image. As for the parameters, the size of the Gaussian filter was set to 1.5, the low threshold was set to 5, and the high threshold was set to 40. Figure 6 shows the locations of the carrier pixels marked in red into which the secret data is to be concealed. On the other hand, Table 1 shows the coordinates of the carrier pixels along with some statistics. Figure 7 is the carrier image after the covering process has been completed.

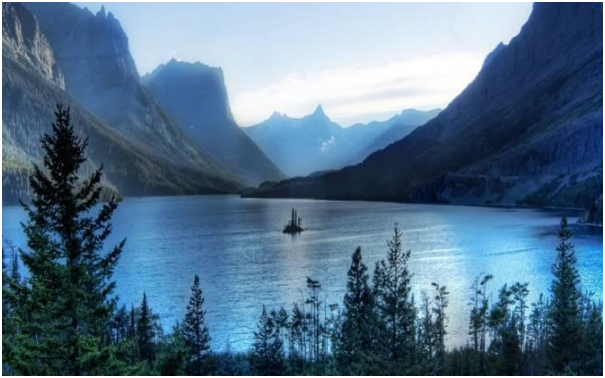

**Fig 4:24-bit Original Input Image**

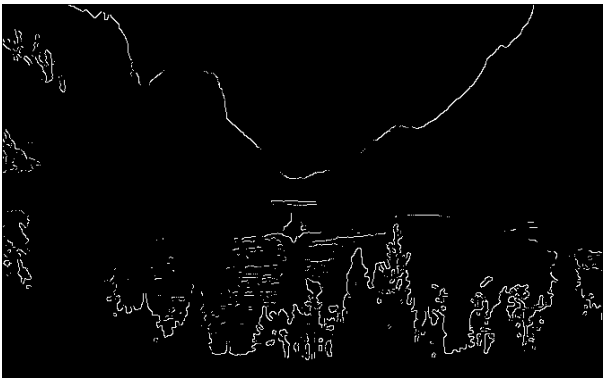

**Fig 5: Results of applying Canny Edge Detection Algorithm**

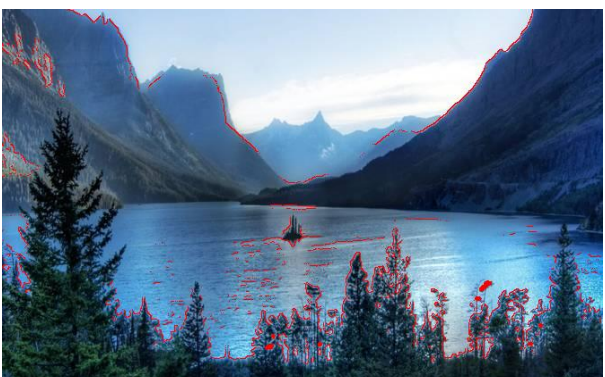

**Fig 6: The red pixels are the carrier pixels**

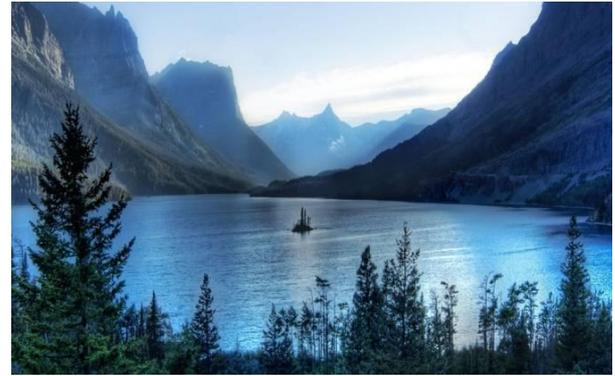

**Fig7: Carrier Image after hiding the secret data into it**

**Table 1. Coordinates of the Carrier Pixels**

| Total Number of pixels in the image | 670x419=280730 pixels |
|---|---|
| Number of pixels that makes up the edges (carrier pixels) | 5630 pixels |
| Hiding capacity for this particular example | 5630 * 3 color channels = 16890 * 3 bits = 50670 bits = 6333 bytes = 6.1 KB |
| Some the carrier pixels coordinates | (024,294) ; (025,294) ; (165,294) ; (166,294) ; (250,294) ; (251,294) ; (252,294) ; (270,294) ; (387,294) ; (388,294) ; (403,294) ; (404,294) ; (415,294) ; (419,294) ; (422,294) ; (423,294) ; (427,294) ; (430,294) ; (431,294) ; (439,294) ; (610,294) ; (657,294) ; (658,294) ; (659,294) ; (660,294) ; (661,294) ; (018,295) ; (025,295) ; (027,295) ; (125,295) ; (249,295) ; (303,295) ; (347,295) ; (387,295) ; (404,295) ; (405,295) ; (406,295) ; (415,295) ; (419,295) ; (423,295) ; (424,295) ; (427,295) ; ………………………….; (339,397) ; (340,397) ; |

## 6. CONCLUSIONS & FUTURE WORK

This paper proposed a steganography technique for hiding digital data into digital images. The technique is based on using Canny edge detection algorithm to pin point the pixels that are part of the objects' boundaries in the image. These pixels are to store the secret data into the three LSBs of its color channels. As for the advantages, the proposed technique is imperceptible as it only uses three LSBs to conceal the secret data into the pixels of the detected edges. Additionally, the proposed technique does not scatter bits of the secret data over all the pixels of the carrier image but only over certain regions that make up the detected edges. A second advantage is theirrecoverability of the hidden data. It is byhiding the secret data using a specific pattern denoted by the Canny algorithm in addition to parameterizing the algorithm allowing the communicating parties to alter the effects and the results of the algorithm, different outputs can be generated for the same input image and secret data. All in all, discovering the inner-workings of the proposed steganography algorithm would be relatively intricate and ambiguous, making the job of eavesdroppers and steganalysts a nightmare, and misguiding them from the exact location of the covert data.

As future work, the proposed technique is to be optimized for other types of digital files such as video files. Moreover, investigating how other image processingtechniquessuch as brightness and contrast adjustment can be exploited in steganography so as to give the communicating parties more options to control their secret communication.

## 7. ACKNOWLEDGMENTS

This research was funded by the Lebanese Association for Computational Sciences (LACSC), Beirut, Lebanon, under the "Stealthy Steganography Research Project – SSRP2012".

## 8. REFERENCES

[1] R. Anderson, F. Petitcolas, "On the limits of steganography," IEEE Journal on Selected Areas in Communications, vol. 16, 1998.

[2] W. Bender, D. Gruhl, N. Morimoto, A. Lu, "Techniques for data hiding IBM Systems Journal", vol. 35, no 3, pp. 313-336, 1996.

[3] Peter Wayner, "Disappearing cryptography: information hiding: steganography & watermarking", 3rd Edition, Morgan Kaufmann Publishers, 2009.

[4] Fabien A. P. Petitcolas, Ross J. Anderson and Markus G.Kuhn, "Information Hiding - A Survey", Proceedings of the IEEE, special issue on protection of multimedia content, vol. 87, no.7, pp.1062-1078, 1999.

[5] Tovée, Martin J., "An introduction to the visual system", Cambridge University Press, 2008.

[6] B. Pfitzmann, "Information hiding terminology", in Information Hiding, First International Workshop, vol. 1174, pp. 347–350, Springer, 1996.

[7] Eric Cole, "Hiding in Plain Sight: Steganography and the Art of Covert Communication", Wiley Publishing, 2003.

[8] J. R. Smith and B. O. Comisky, "Modulation and information hiding in images," in information hiding, first international workshop, Germany: Springer-Verlag, vol. 1174, pp. 207–226, 1996.

[9] Fabien A. P. Petitcolas, Ross J. Anderson and Markus G.Kuhn, "Information Hiding - A Survey", Proceedings of the IEEE, special issue on protection of multimedia content, vol. 87, no.7, pp.1062-1078, 1999.

[10] Tovée, Martin J., "An introduction to the visual system", Cambridge University Press, 2008.

[11] W. Bender, D. Gruhl, N. Morimoto, and A. Lu, "Techniques for data hiding", IBM Systems Journal, vol. 35, no. 3-4, pp. 313-336, 1996.

[12] Johnson, N. F. and Jajodia, S., "Exploring steganography: Seeing the unseen", Computer Journal, vol. 31, no.2, pp.26–34, 1998.

[13] T. Zhang and X. Ping, "A Fast and Effective Steganalytic Technique against JSteg-like Algorithms", Proceedings of the 8th ACM Symposium, Applied Computing, ACM Press, 2003.

[14] Eiji Kawaguchi and Richard O. Eason, "Principle and applications of BPCS-Steganography", Proceedings of SPIE: Multimedia Systems and Applications, vol.35, no.28, pp.464-473, 1998.

[15] T. Lindeberg (1998) "Edge detection and ridge detection with automatic scale selection", International Journal of Computer Vision, 30, 2, pages 117--154.

[16] H.G. Barrow and J.M. Tenenbaum (1981) "Interpreting line drawings as three-dimensional surfaces", Artificial Intelligence, vol 17, issues 1-3, pages 75-116.

[17] Lindeberg, Tony (2001), "Edge detection", in Hazewinkel, Michiel, Encyclopedia of Mathematics, Springer, ISBN 978-1-55608-010-4

[18] Shapiro L.G. & Stockman G.C. (2001) Computer Vision. London etc.: Prentice Hall, Page 326.

[19] Canny, J., A Computational Approach To Edge Detection, IEEE Trans. Pattern Analysis and Machine Intelligence, 8(6):679–698, 1986.

[20] Charles Petzold, "Programming Microsoft Windows with C#", Microsoft Press, 2002.

[21] AForge.NET Framework, URL: http://www.aforgenet.com/framework/, Retrieved October, 27, 2012.